\documentclass[twocolumn]{aa}
\usepackage{epsfig}
\def\simlt{\ \raise -2.truept\hbox{\rlap{\hbox{$\sim$}}\raise5.truept   %
\hbox{$<$}\ }}
\def\simgt{\ \raise -2.truept\hbox{\rlap{\hbox{$\sim$}}\raise5.truept   %
\hbox{$>$}\ }}                                                          %

\def\be{\begin{equation}}
\def\ee{\end{equation}}
\def\newline{\hfil\break}

\def\la{\mathrel{\hbox{\rlap{\hbox{\lower4pt\hbox{$\sim$}}}\hbox{$<$}}}}
\def\ga{\mathrel{\hbox{\rlap{\hbox{\lower4pt\hbox{$\sim$}}}\hbox{$>$}}}}

\def\es{{\rm 1ES0657-556}~}
\begin{document}
\title{Direct probes of Dark Matter in the cluster 1ES0657-556 through microwave
observations}
   \author{S. Colafrancesco\inst{1},
   P. de Bernardis\inst{2},  S. Masi\inst{2}, G. Polenta\inst{2} and P.
Ullio\inst{3}}
   \offprints{S. Colafrancesco}
\institute{   INAF - Osservatorio Astronomico di Roma
              via Frascati 33, I-00040 Monteporzio, Italy.
              Email: cola@mporzio.astro.it
 \and
              Dipartimento di Fisica, Universit\`a Roma 1, P.le A. Moro 2,
Roma, Italy
 \and
              Scuola Internazionale Superiore di Studi Avanzati,
              Via Beirut 2-4, I-34014 Trieste, Italy
             }
\date{Received 16 October 2006 / Accepted 25 January 2007 }
\authorrunning {S. Colafrancesco et al.}
\titlerunning {SZ effect from 1ES0657-556 cluster}
\abstract
   {}
   {The cluster \es is an ideal astrophysical laboratory to study the
   distribution and the nature of Dark Matter because this last component is spatially
   separated from the intracluster gas.
   We show that microwave observations can provide crucial probes of Dark Matter in this
   system.}
  {We calculate the expected SZ effect from Dark Matter annihilation in the main mass
  concentrations of the cluster 1ES0657-556, and we estimate the sources of contamination,
  confusion and bias to asses its significance.}
   {We find that SZ observations at $\nu \approx 223$ GHz can resolve both spatially and
   spectrally the SZ$_{DM}$ signal and isolate it from the other SZ signals,
   and mainly from the thermal SZ effect which is null at $ \nu \sim 220-223$ GHz
   for the case of \es. We conclude that SZ observations with $\simlt$ arcmin resolution
   and $\simlt \mu$K sensitivity of \es are crucial, and maybe unique, to find direct
   astrophysical probes of the existence and of the nature of Dark Matter,
   or to set strong experimental limits.}
  {}

 \maketitle
\section{Introduction}
Dark Matter (DM) annihilations in the halo of galaxies and galaxy clusters have crucial
astrophysical implications. In fact, if DM is constituted by weakly interacting massive
particles (for which the leading candidate is the lightest supersymmetric particle,
plausibly the neutralino $\chi$), their annihilation produces secondary particles (e.g.,
neutral and charged pions, secondary electrons and protons, neutrinos) that give rise to
various astrophysical signals. These are, among others, observable fluxes of positrons,
antiprotons, gamma rays, neutrinos, as well as signals due to secondary electrons which
cover the whole e.m. spectrum (see, e.g., Colafrancesco et al. 2006 for details):
synchrotron radio emission (in the intra-cluster magnetic field), bremsstrahlung emission
(if there is co-spatial intra-cluster gas), inverse Compton emission due to the
up-scattering of CMB photons and hence a specific SZ effect (as first noticed and derived
by Colafrancesco 2004).
The spatial and spectral intensity of the astrophysical signals coming from $\chi \chi$
annihilation is expected, however, to be confused or even overcome by other astrophysical
signals originating from the intracluster (IC) gas and/or from the relativistic plasmas
present in the cluster atmospheres, especially when all these components are co-spatially
distributed with the DM component. This situation occurs in most galaxy clusters (see
discussion by Colafrancesco et al. 2006 for the case of Coma).\\
An ideal system to detect DM annihilation signals would, therefore, be a cluster with a
clear spatial separation between the various matter components. This is, indeed, the case
of the cluster 1ES0657-556 where the spatial distribution of DM is clearly offset w.r.t.
that of the IC gas (Clowe et al. 2006). The two baryonic clumps of hot gas emit X-rays by
(thermal) bremsstrahlung, as observed by Chandra (Markevitch et al. 2002, 2004). The
shock observed in the western-most region of the cluster (Markevitch et al. 2002) might
be the site of high energy emission from particles accelerated at the shock.
Hard X-ray emission from the direction of 1ES0657-556 has been marginally detected by
Rossi-XTE (Petrosian et al. 2006) but its angular resolution is not sufficient to give
any information on the spatial distribution of this emission. No gamma-ray emission has
been detected from this system with EGRET.
The extended radio halo associated to this cluster (Liang et al. 2000) has a surface
brightness slightly elongated along the direction of the two X-ray clumps, but without
clear signatures of radio-brightness enhancements at the DM clump locations.
Finally, the SZ maps of 1ES0657-556 obtained with ACBAR (with $\sim 4.5$ arcmin FWHM
resolution, Gomez et al. 2003) are quite smooth and regular with no evidence of
enhancement at both X-ray and/or DM clump locations.
In this Letter, we will compute the specific feature of the SZ effect (herefater SZE)
produced by DM annihilation, SZ$_{\rm DM}$, in the cluster \es and we will show that it
is possible to detect such SZ$_{\rm DM}$ signal with a specific observational strategy.
The relevant physical quantities are calculated using $H_0 = 70$ km s$^{-1}$ Mpc$^{-1}$
and a flat, $\Lambda$CDM ($\Omega_{\rm m} = 0.3, \Omega_{\Lambda}=0.7$) cosmological
model.
\section{The complex SZ effect in \es}
The various SZ signals expected from the subsystems of the cluster \es are:
i) the SZ$_{\rm DM}$ effect, which is expected to be located at the two DM clumps;
ii) the thermal SZ effect (SZ$_{\rm th}$) which is expected to be located at the two
X-ray clumps.
We will compute in the following these two sources of SZE and we will also discuss the
possible sources of contamination, bias and confusion.
The general expression for the SZE  which is valid in the Thomson limit for a generic
electron population in the relativistic limit and includes also the effects of multiple
scatterings and the combination with other electron population in the cluster atmospheres
has been derived by Colafrancesco et al. (2003). This approach is the one that will be
used for the derivation of the SZ$_{\rm DM}$ effect induced by the secondary electrons
produced by $\chi \chi$ annihilation (see the original derivation by Colafrancesco 2004)
as well as for the SZ$_{th}$ produced by the hot X-ray emitting gas.
According to these results, the spectral distortion of the CMB spectrum induced by a
population of electron with momentum distribution $f_e(p)$ can be written as
 \begin{equation}
\Delta I(x)=2\frac{(k_{\rm B} T_0)^3}{(hc)^2} ~\tilde{y} ~\tilde{g}(x) ~,
\end{equation}
where $T_0$ is the CMB temperature and
\begin{equation}
\tilde{y} =\frac{\sigma_T}{m_{\rm e} c^2}\int P_{\rm e} d\ell ~,
\end{equation}
in terms of the pressure $P_{e}$ contributed by the specific electron population.
The spectral function $\tilde{g}(x)$, with $x \equiv h \nu / k_{\rm B} T_0$, can be
written as
\begin{equation}
\label{gnontermesatta} \tilde{g}(x)=\frac{m_{\rm e} c^2}{\langle \epsilon \rangle}
\left\{ \frac{1}{\tau} \left[\int_{-\infty}^{+\infty} i_0(xe^{-s}) P(s) ds- i_0(x)\right]
\right\}
\end{equation}
in terms of the photon redistribution function $P(s)$ and of $i_0(x) = 2 (k_{\rm B}
T_0)^3 / (h c)^2 \cdot x^3/(e^x -1)$.
Here $\tau = \int d \ell n_e$ is the optical depth of the electrons with number density
$n_e$, and
$ \langle \epsilon \rangle \equiv \frac{\sigma_{\rm T}}{\tau}\int P_{\rm e}
d\ell
 = \int_0^\infty dp f_{\rm e}(p) \frac{1}{3} p v(p) m_{\rm e} c
$
is the average energy of the electronic plasma.
The photon redistribution function $P(s)= \int dp f_{\rm e}(p) P_{\rm s}(s;p)$ with $s =
\ln(\nu'/\nu)$, in terms of the CMB photon frequency increase factor $\nu' / \nu$,
depends on the electron momentum  distribution $f_{\rm e}(p)$, where the momentum $p$ is
normalized to $m_e c$.
The CMB temperature change produced by the SZE is finally given by
 \be
{\Delta T \over T_{0}} = {(e^x-1)^2 \over x^4 e^x} {\Delta I \over I_0}
\,.
 \label{eq.deltat}
 \ee
The specific SZ$_{DM}$ and SZ$_{th}$ effects for the various electronic components in the
cluster \es are computed following the approach previously described.

\subsection{The SZ$_{DM}$ effect in the cluster \es}
\begin{figure}[!h]
\begin{center}
\vbox{
 \hbox{
 \epsfig{file=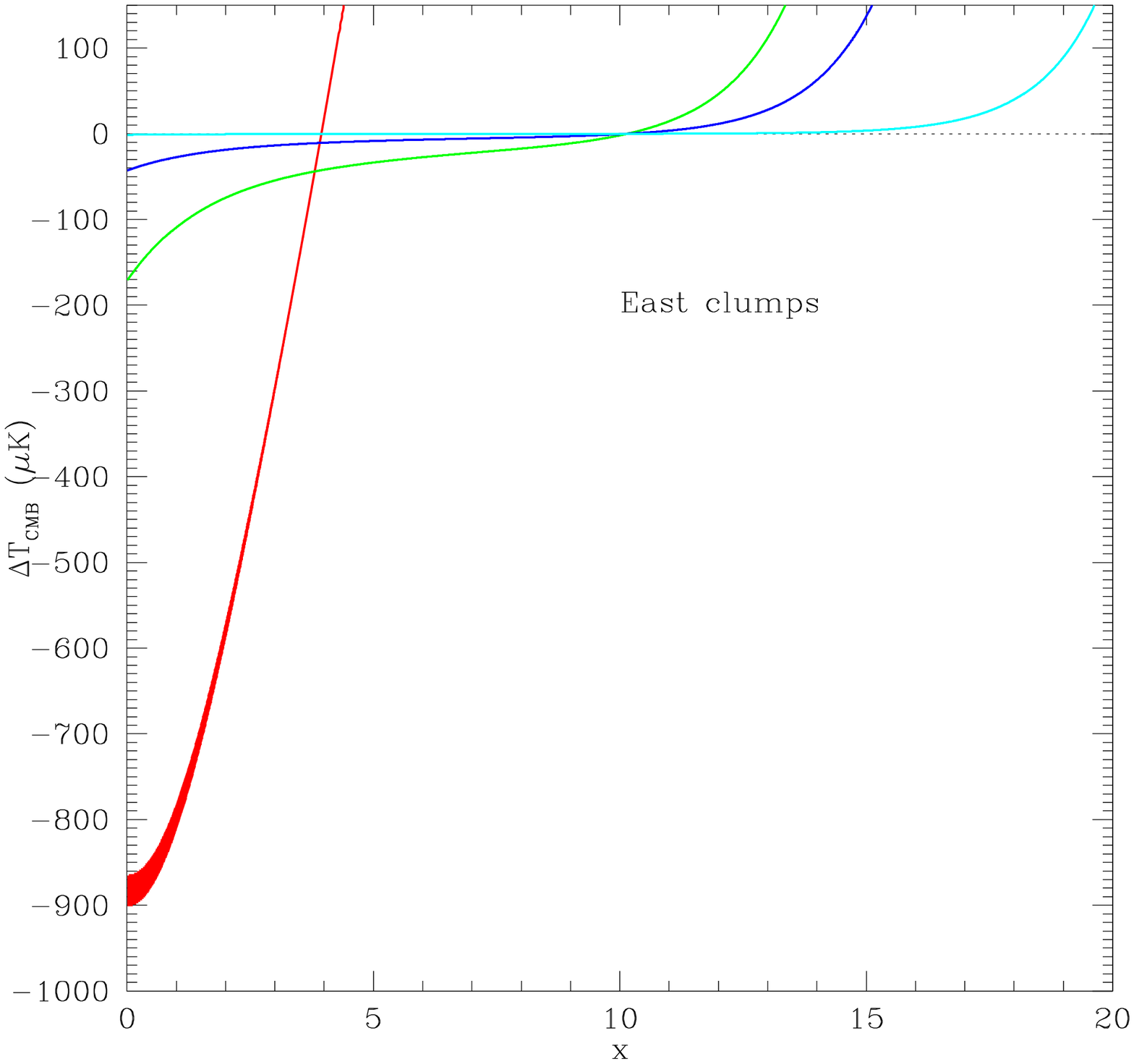,height=5.8cm,width=4.5cm,angle=0.0}
 \epsfig{file=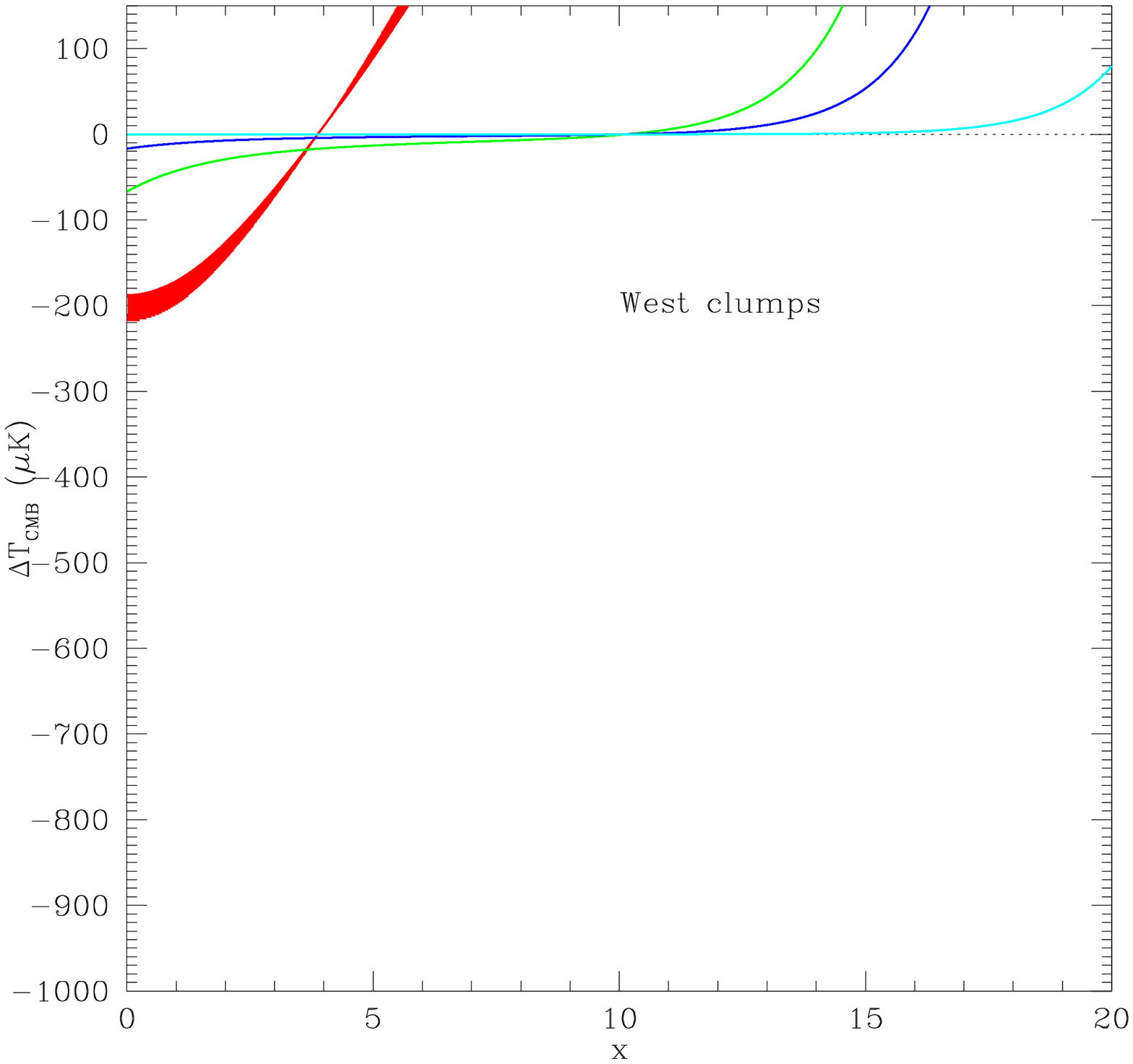,height=5.8cm,width=4.5cm,angle=0.0}
}
 \hbox{
 \epsfig{file=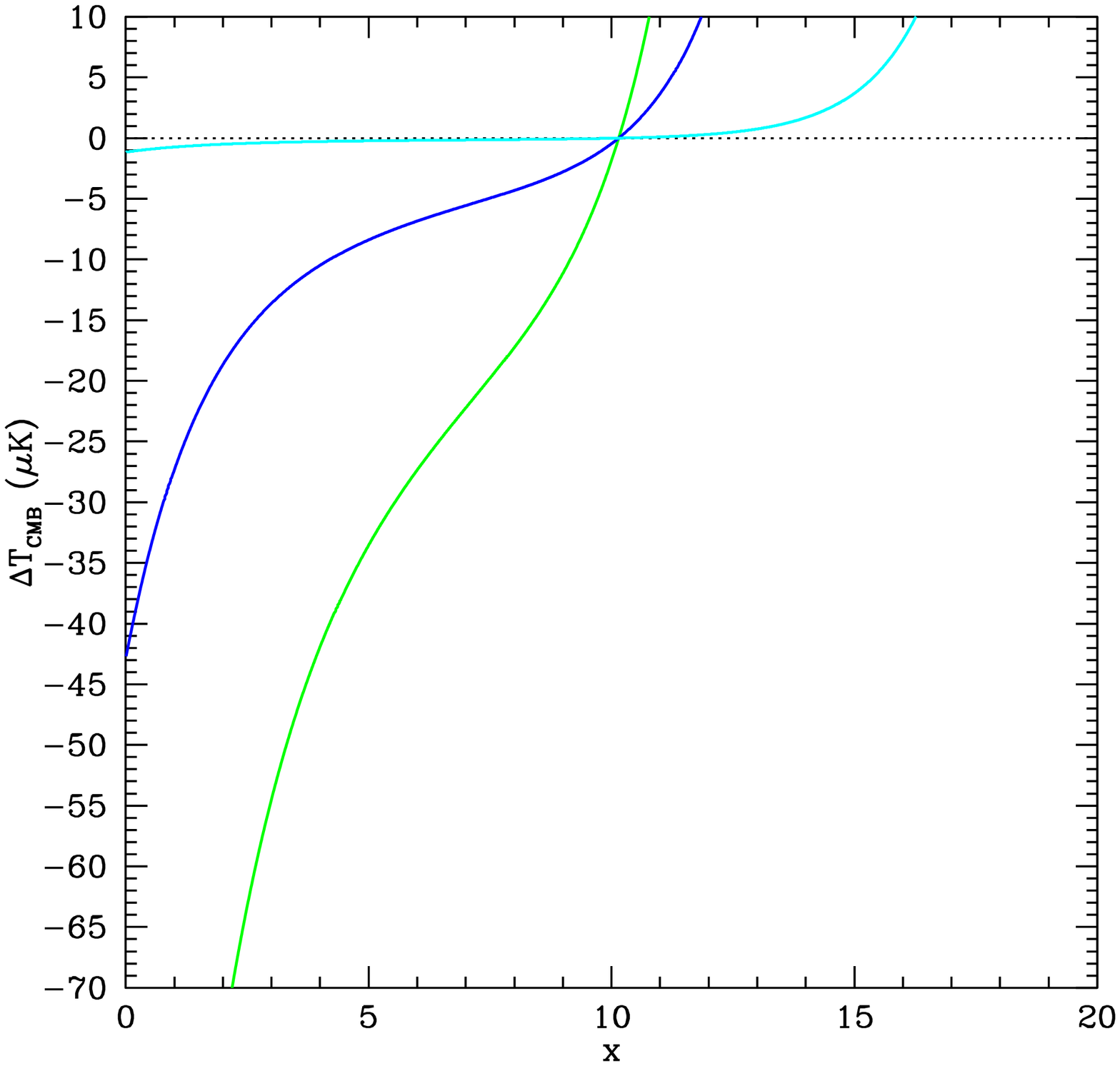,height=5.8cm,width=4.5cm,angle=0.0}
 \epsfig{file=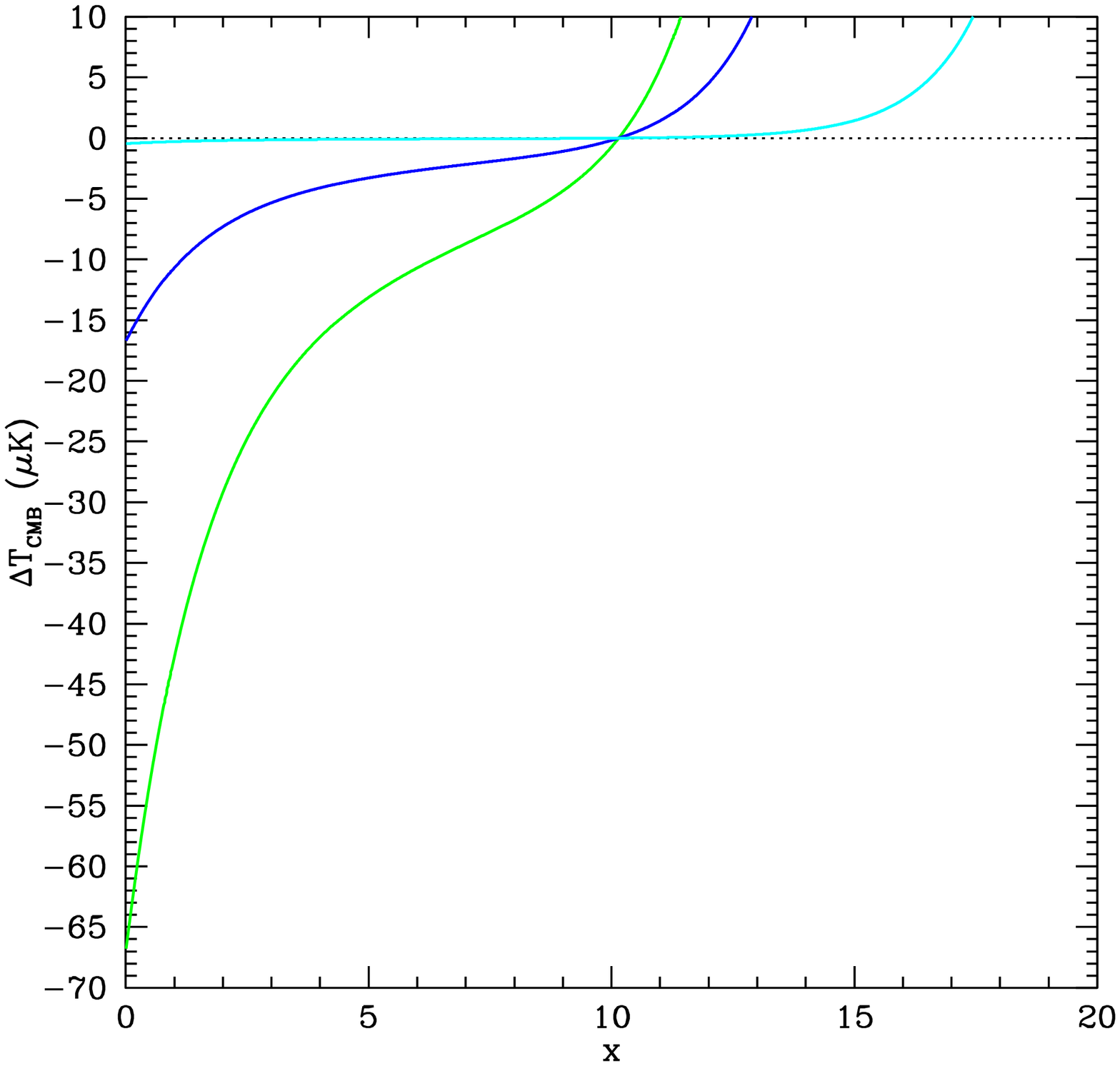,height=5.8cm,width=4.5cm,angle=0.0}
 }
 }
\end{center}
  \caption{\footnotesize{The CMB temperature decrement produced by the
East (left panels)  and West (right panels) clumps of gas (red shaded areas which take
into account the effect of IC gas temperature uncertainties) and DM with neutralino mass
$M_{\chi}=20$ GeV (green), $M_{\chi}=40$ GeV (blue) and $M_{\chi}=81$ GeV (cyan).  The
lower panels show a zoom of the SZ$_{DM}$ effect only.
  }}
  \label{fig.sz_dm}
\end{figure}

The calculation of the secondary electron spectrum from $\chi \chi$ annihilation in
galaxy clusters has been already presented in details by Colafrancesco \& Mele (2001) and
Colafrancesco et al. (2006), and here we will only recall the relevant steps necessary
for the present purpouses.\\
We assume, for simplicity, a spherical DM halo model for each DM clump of the cluster
\es, as indicated by the lensing maps derived by Clowe et al. (2006), with DM density
profile given by $ g(x) = x^{-\eta} (1+ x)^{\eta - \xi}$, with $x \equiv r/r_s$. Values
$\eta = 1$ and $\xi = 3$ reproduce the Navarro, Frenk \& White (1997) density profile.\\
The neutralino number density profiles $n_{\chi}(E,r) = n_{\chi,0}(E) g(r)$ of the two DM
clumps have been calculated following the approach described in Colafrancesco et al.
(2006) with a NFW DM density profile and the following structure parameters: $M_{vir}=
10^{15} M_{\odot}$, $R_{vir} = 1.97$ Mpc and $c_{vir}= R_{vir}/r_s = 5.66$ (for the
larger East DM clump); $M_{vir} = 6.25 \cdot 10^{13} M_{\odot}$, $R_{vir} = 0.784$ Mpc
and $c_{vir}= 7.56$ (for the ''bullet'' West DM clump).\\
The $\chi$ annihilation rate in the DM clumps is $ R = n_{\chi}(r) \langle \sigma v
\rangle_0 ~,$ where $\langle \sigma v \rangle_0$ is the $\chi \chi$ annihilation cross
section averaged over a thermal velocity distribution at freeze-out temperature.
The range of neutralino masses and pair annihilation cross sections in the most general
supersymmetric DM setup is extremely wide (see discussion in Colafrancesco et al. 2006,
2007).
We consider here, specifically, the neutralino models worked out in Colafrancesco et al.
(2006) with $M_{\chi}= 20, 40$ and $81$ GeV and with their specific values of $\langle
\sigma v \rangle_0$.\\
The electron source functions $Q_{\rm e}(E,r) \propto \langle\sigma v\rangle_0
n^2_{\chi}(E,r)$ for the specific neutralino model considered here have been derived in
Colafrancesco et al. (2006) and the time evolution of the electron spectrum is given by
the equation
 \be
{\partial n_{\rm e} (E,r)\over \partial t} - {\partial  \over \partial E} \bigg[ n_{\rm
e} (E,r) b(E)\bigg] = Q_{\rm e}(E,r) \, ,
 \label{eq.diffusion}
 \ee
where spatial diffusion can be safely neglected in cluster-size DM clumps (Colafrancesco
et al. 2006).
The function
\begin{eqnarray}
 b(E) & = & b_{IC}^0 \left(\frac{E}{\rm{GeV}}\right)^{2} + b_{syn}^0 B_\mu^2
\left(\frac{E}{\rm{GeV}}\right)^{2}
 \nonumber \\
 & & + b_{Coul}^0 n_{th} \left(1+\log(\gamma / n_{th})/75 \right)  \nonumber \\
 & & + b_{brem}^0 n_{th} \left(\log(\gamma / n_{th})+0.36 \right)
 \end{eqnarray}
gives the energy loss per unit time at energy $E$ where $n_{th}$ is the mean number
density of thermal electrons in $\rm{cm}^{-3}$, $\gamma \equiv E/m_e c^2$ and $b_{IC}^0
\simeq 0.25$, $b_{syn}^0 \simeq 0.0254$, $b_{Coul}^0 \simeq 6.13$, $b_{brem}^0 \simeq
1.51$, all in units of $10^{-16}\; \rm{GeV}\, \rm{s}^{-1}$.
The equilibrium spectrum $n_e(E,r)$ obtained solving eq.(\ref{eq.diffusion}) allows  to
calculate the SZ$_{DM}$ effect.
Fig.\ref{fig.sz_dm} shows the CMB temperature change $\Delta T$ evaluated at the centers
of the two DM clumps for different values of $M_{\chi}$.
The SZ$_{DM}$ signals are overcome by the SZ$_{th}$ signals at low ($\nu \simlt 200$ GHz)
and high ($\nu \simgt 230$ GHz) frequencies, while they dominate in the frequency range
where the zero of SZ$_{th}$ is found, i.e. at $\nu \approx 223$ GHz for the East clump
and at $\nu \approx 219$ GHz for the West clump. At such frequencies the SZ$_{DM}$
temperature decrement takes values $\approx -21.1, -10.6, -0.3$ $\mu$K (East clump) and
values $\approx -8.4, -4.2, -0.1$ $\mu$K (West clump) for $M_{\chi} = 20, 40, 81$ GeV,
respectively.
\subsection{The SZ$_{th}$ effect in the cluster \es}
The SZ$_{th}$ from the two X-ray clumps of the cluster \es is also computed from the
general approach delineated in eqs.(1-4). Following Markevitch et al. (2002, 2004) and
Tucker et al. (1998), we use the simplifying assumption that the IC gas distribution can
be described by isothermal spheres fitted by a $\beta$-profile.
For the East X-ray clump we adopt $kT_e = 14$ keV (Markevitch et al. 2002, 2004) and a
$\beta$-profile with $r_c=257$ kpc, $\beta=0.62$ and a central IC gas density
$n_{th,0}=6.71 \cdot 10^{-3}$ cm$^{-3}$ (Tucker et al. 1998).
For the West X-ray clump we adopt $kT_e = 6$ keV  and a $\beta$-profile with $r_c=125$
kpc, $\beta=0.7$ and a central IC gas density $n_{th,0}=2 \cdot 10^{-2}$ cm$^{-3}$,
neglecting the additional central power-law density profile induced by the presence of
the shock (Markevitch et al. 2002, 2004). Thus, SZ$_{th}$ from this last X-ray clump
should be considered as slightly underestimated.\\
The possible non-equilibrium and non-thermal effects induced by the merging shock are not
crucial for the aims of this paper since the shock is located outside the two X-ray
clumps and at a large angular distance from the two DM clumps (see Markevitch et al.
2002, 2004). Possible non-thermal SZE produced by shocks is, however, expected to be at
most at a level of a few $\mu$K (see Colafrancesco et al. 2003), and therefore it could
be, at most, a source of bias for the thermal SZE at the location of the X-ray clumps.
The SZ$_{th}$ temperature change for the two X-ray clumps is also shown in
Fig.\ref{fig.sz_dm}.
A major difference between the SZ$_{th}$ and SZ$_{DM}$ spectral functions is the
different position of the zero of the SZE which is moved to higher frequencies in the
case of the DM produced electrons with respect to the case of the thermal distribution.
As a consequence, the SZ$_{\rm DM}$ effect appears as a negative contribution to the
overall SZE at all the frequencies which are relevant for the SZ experiments, $x \sim 0.5
- 10$.

\subsection{Contamination, bias and confusion}
 \label{sec.confusion}
\begin{figure}[!h]
\begin{center}
 \hbox{
 \hspace{-3.cm}
 \vbox{
 \epsfig{file=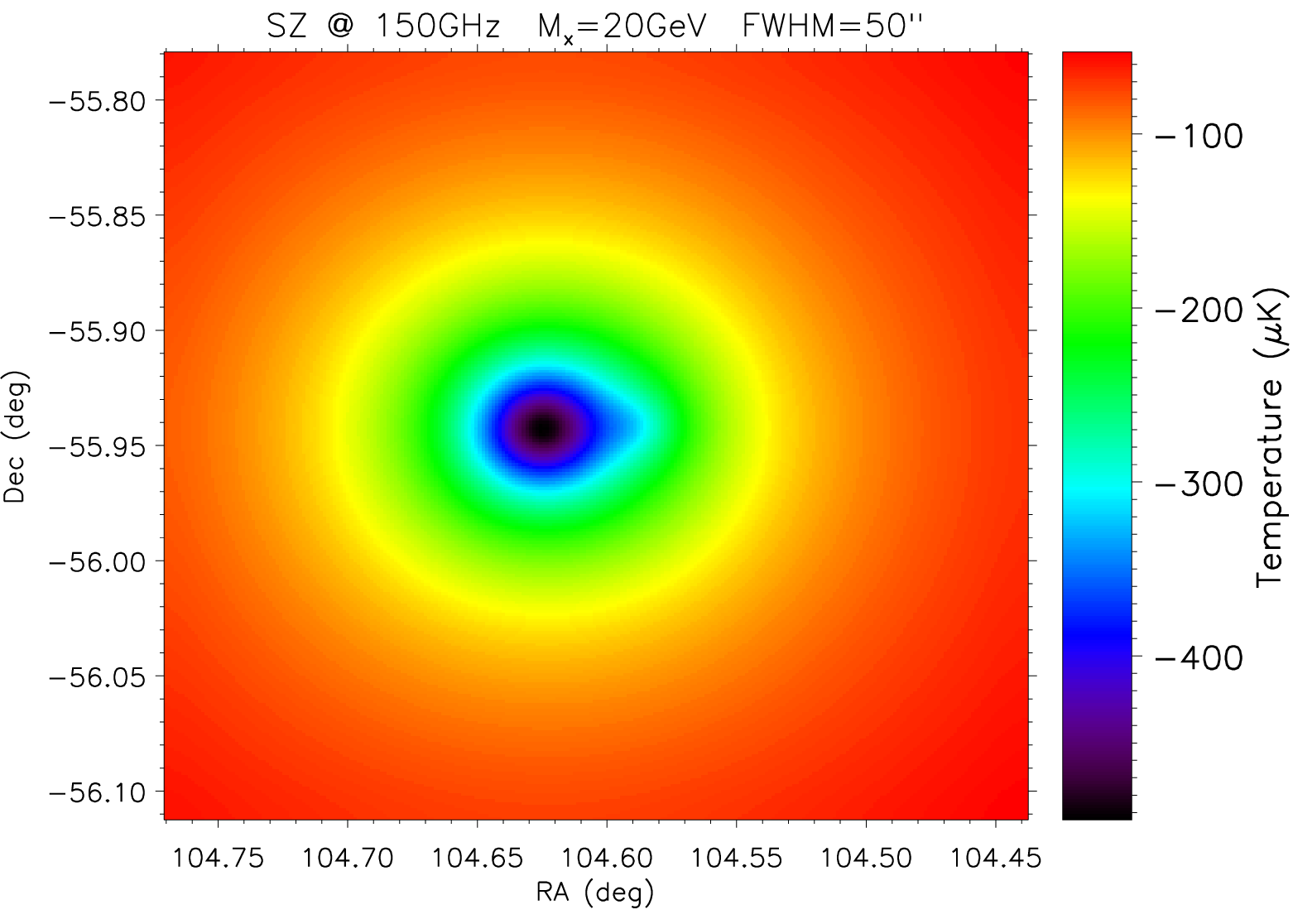,width=5cm,angle=0.0}
 \epsfig{file=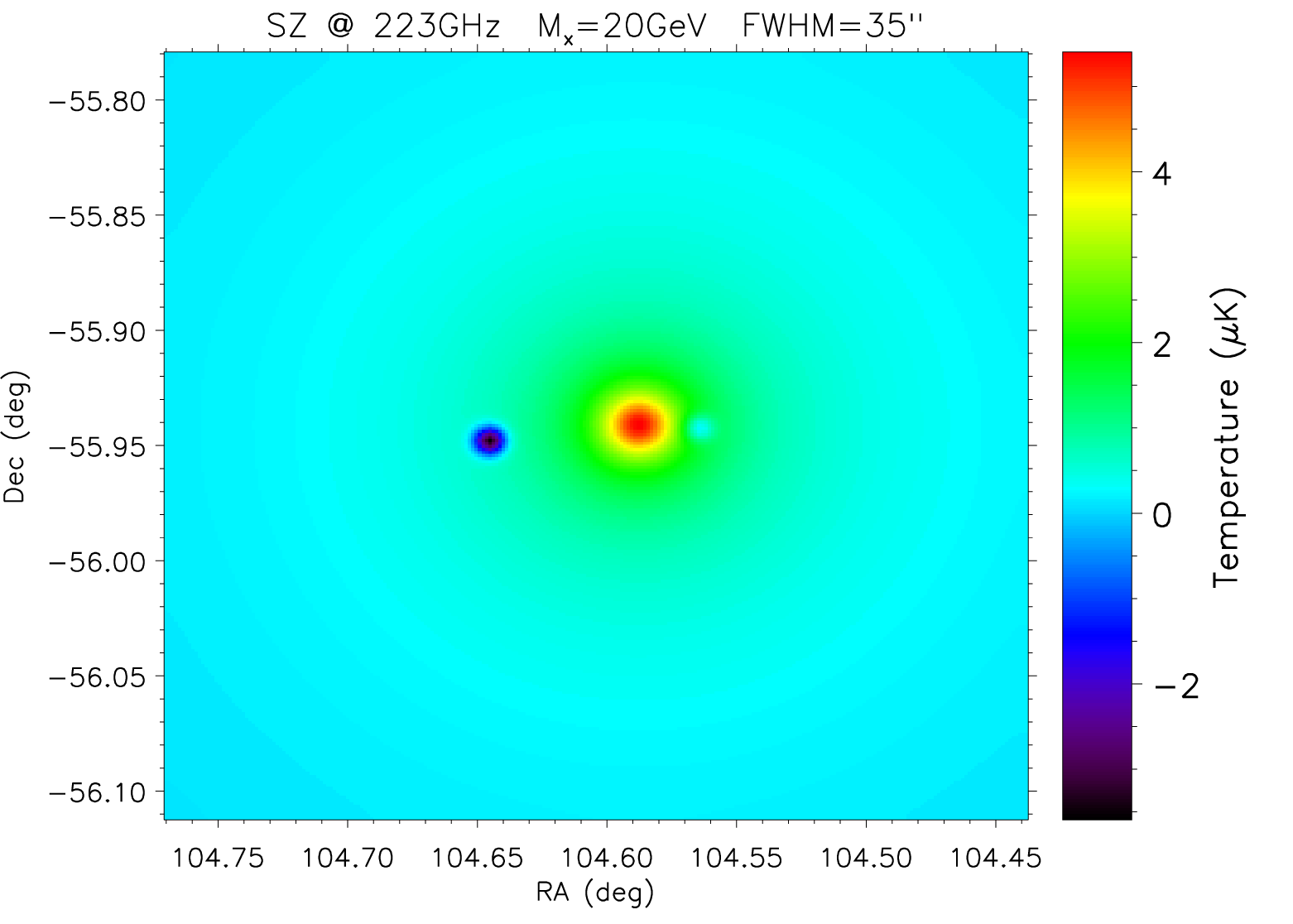,width=5cm,angle=0.0}
 \epsfig{file=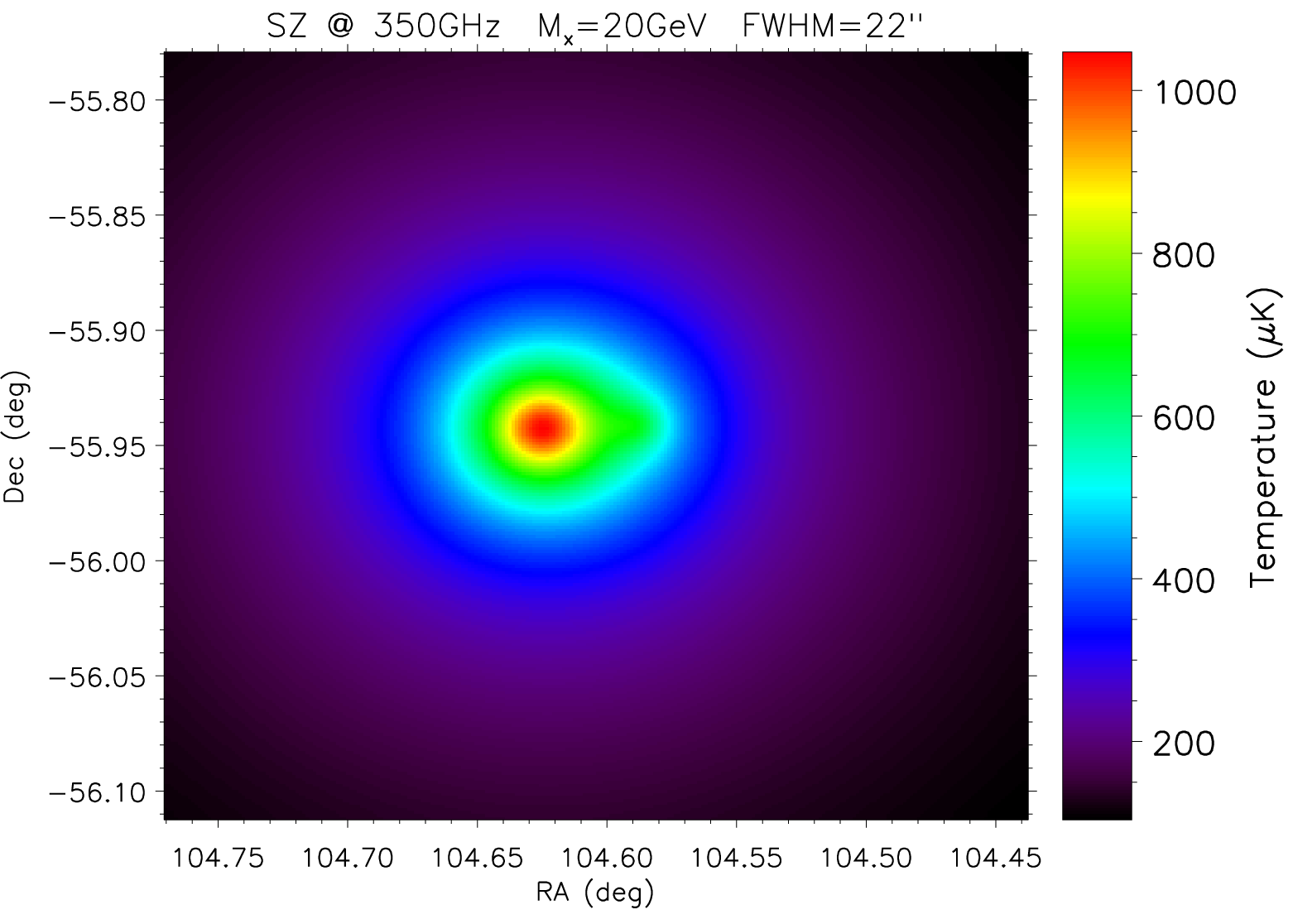,width=5cm,angle=0.0}
 }
 \hspace{-4.cm}
 \vbox{
 \epsfig{file=6569fig6.ps,width=5cm,angle=0.0}
 \epsfig{file=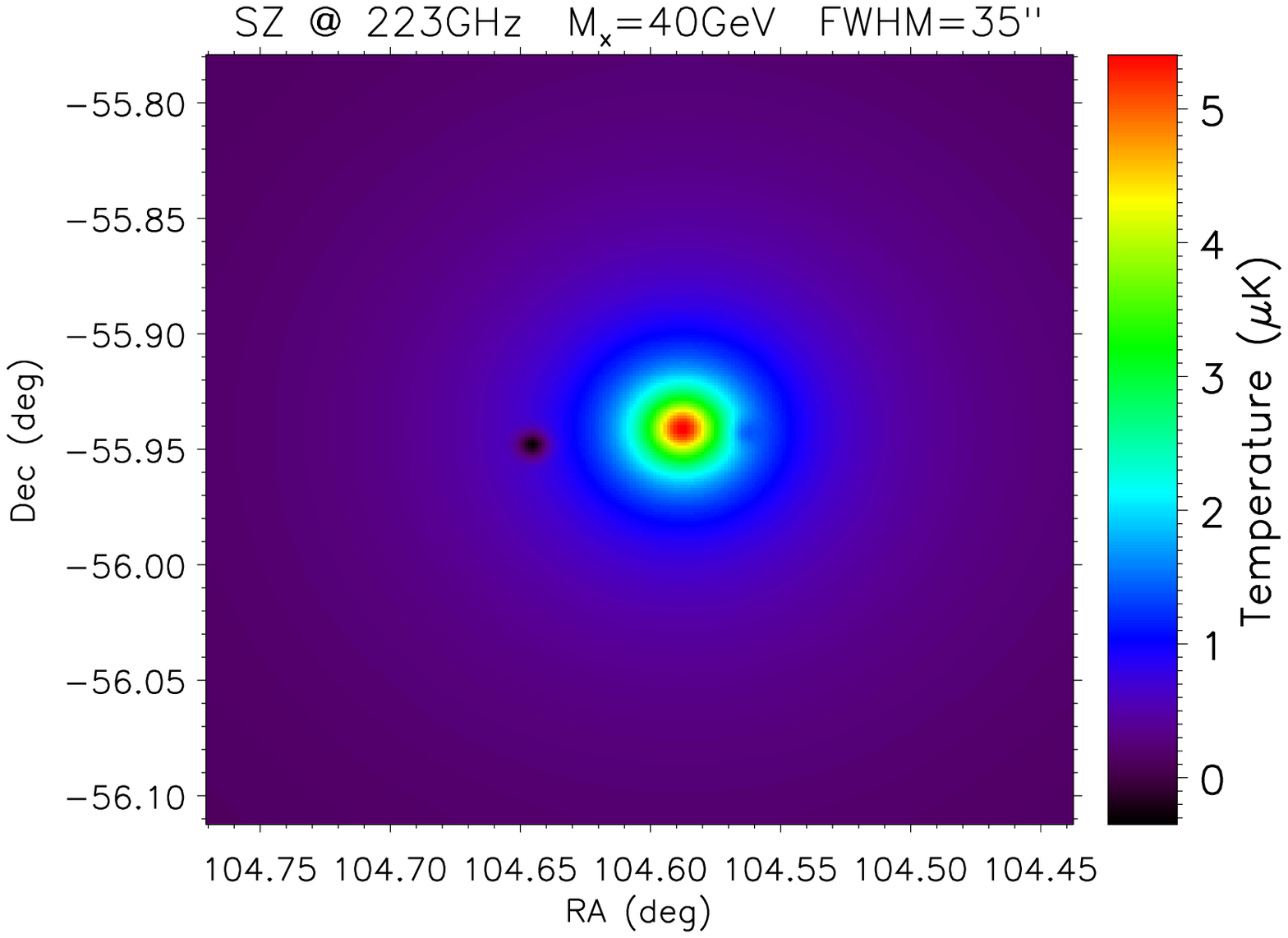,width=5cm,angle=0.0}
 \epsfig{file=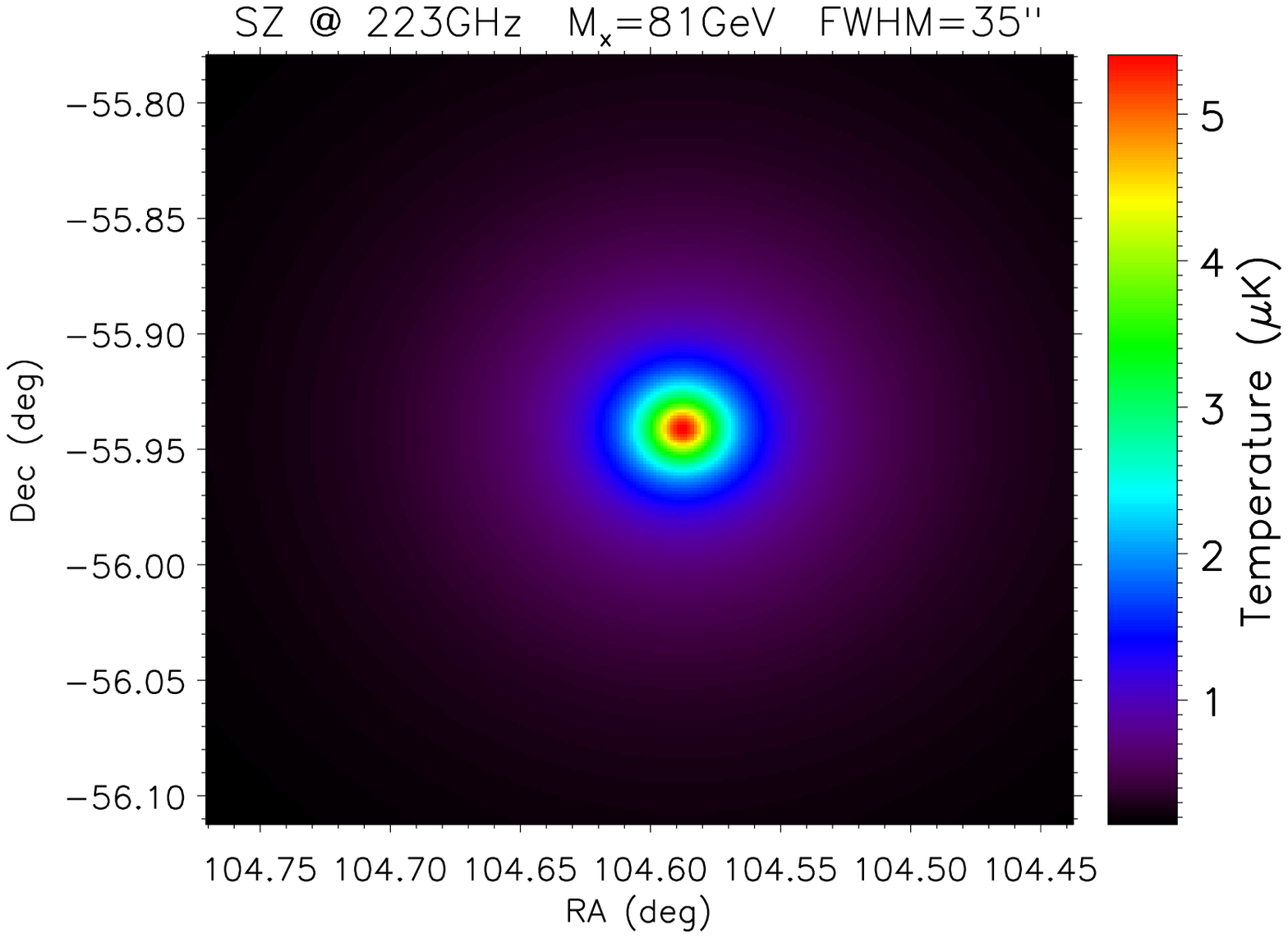,width=5cm,angle=0.0}
 }
 }
  \caption{\footnotesize{Left. Simulated SZ maps of the cluster \es as observable with
  the SPT at $\nu = 150$ GHz (upper panel), $\nu = 223$ GHz (mid panel), $\nu = 350$ GHz
  (lower panel). A value $M_{\chi}=20$ GeV has been adopted here.
  Right. Simulated SZ maps of the cluster \es as observable with
  the SPT at $\nu = 223$ GHz for three different $\chi$ masses:
  $M_{\chi}=20$ GeV (upper panel), 40 GeV (mid panel) and 81 GeV (lower panel).
  }}
  \label{fig.sz_maps}
\end{center}
\end{figure}

Possible sources of confusion for the microwave search for DM  in \es are: CMB
anisotropy, emission of the IC medium, emission of galaxies and AGNs along the line of
sight, thermal and kinematic SZE from the X-ray clumps. For the estimates of the
foreground contamination we consider as peaks of the DM distribution producing SZE the
regions included in the $\kappa = 0.3$ contour of Fig.1 in Clowe et al. (2006). These are
roughly 1.3$^\prime$ in diameter, and are separated by $\sim 4.8^\prime$.\\
The measurement we propose must be carried out with an instrument with angular resolution
of $\sim$ 1$^\prime$  FWHM, able to resolve the two SZ$_{DM}$ peaks, and to distinguish
them from the thermal and kinematic SZE of the two X-ray clumps. Larger scales are not
significant for the measurement, and must be removed either in the measurement or in the
data analysis. This will remove most of the CMB anisotropy. In fact, integrating over the
power spectrum of CMB anisotropy best fitting current measurements (including lensing
effects), we find that an angular high-pass filter removing scales $\simgt 3 ^\prime$
will cut multipoles $\ell \simlt 3600 $, leaving a residual rms fluctuation of primary
CMB anisotropy of 0.8 $\mu$K rms. This is further reduced by the finite angular
resolution of the telescope.\\
In the two SZ$_{DM}$ peaks we find 14 galaxies in the area of the East peak and 6
galaxies in the area of the West peak (see Barrena et al. 2002). To obtain an estimate of
the mm-wave emission of these galaxies, we take the mm-wave spectrum of an "average",
"normal" galaxy (like M99) and redshift it to z=0.3 . Then we compute the flux in the
bands centered at 90, 145, 217, 270, 345, 545 GHz and we find the total flux by
multiplying it by the number of galaxies counted in the West and East SZ$_{DM}$ peaks. In
the West peak, we obtain a total flux of 0.2, 0.4, 1.0, 1.6, 3.4, 19 mJy, respectively,
in the bands quoted above. These signals correspond to a CMB anisotropy of 4.0, 4.8, 8.0,
13, 40, and 760 $\mu K$ for a 1$^\prime$ FWHM beam. These are comparable in size to
expected SZ$_{DM}$ signal, at least in the 90, 145, and 217 GHz bands, but are positive
signals, while the SZ$_{DM}$ signal is negative at the same frequencies. The same holds
for AGNs: there is only one known radio source (SUMSS J065837-555718) listed in the NED
database in the area of the East peak, and none in the area of the West peak.
This source has been observed with the MOST radio telescope, and features a power law
SED, with slope $\alpha \sim -0.9$. If the same SED is used to extrapolate to the
frequencies of interest here, the total flux is safely negligible with respect to the
expected SZ$_{DM}$ in all the bands considered here: for a 1$^\prime$ FWHM beam the
signal is below 1 $\mu K$ in the 145, 217, 270 and 345 GHz bands.\\
Unresolved background galaxies produce a flux with positive and negative fluctuations
with respect to the average. For a 1$^\prime$ beam at 220 GHz, we get a rms signal of
$\sim 3 \mu K$, estimated using the model of Lagache et al. (2004).\\
The SZ$_{th}$ emission from the IC gas is offset with respect to the DM peaks, so that
only the tail of the signal is present in the SZ$_{DM}$ peaks.
The SZ$_{th}$ emission is certainly affected by the complex temperature distribution of
the X-ray clumps (see Markevitch et al. 2002) producing fluctuations in the SZ maps
(since $\Delta T \propto \int d\ell n_e T_e $).
This makes the maps at low (150 GHz) and high (350 GHz) frequency shown in Fig.2 as only
indicative of the observable SZ signal.
However, we stress that the SZ maps at $\nu \approx 219-223$ GHz are quite realistic
because at 223 (219) GHz for the East (West) X-ray peak, the SZ$_{th}$ is null, and
temperature uncertainties do not sensitively alter this result (see Fig.\ref{fig.sz_dm}):
this is, henceforth, the privileged band for this measurement.\\
The $(\Delta T/T_0)_{kin}$ at the clump centers is $\simlt 2.7 \cdot 10^{-6} (\tau
/10^{-3})(V_p/10^3 km s^{-1})$ and is maximum at 217 GHz.  Small values of $\Delta
T_{kin}$ are found at the location of the two DM clumps because: i) the optical depth of
the secondary electrons is small $\simlt {\rm a~few~} 10^{-5}$; ii) the optical depth of
the residual IC gas at the DM peak locations is small $\simlt {\rm a~ few~} 10^{-4}$, and
iii) the peculiar velocity of the X-ray clumps and DM clumps along the line of sight is
relatively small (Markevitch 2004), since the major merging event occurs almost
completely on the plane of the sky, i.e. mostly perpendicular to the line of sight. Even
an extreme value $V_p \approx 600$ km/s  (well above the rms value $\approx 300 \pm 80$
km/s found for galaxy clusters, see Giovanelli et al. 1998) would produce a total $\Delta
T_{kin} \sim 0.4$ $\mu$K ($\sim 0.2$ $\mu$K) in the East (West) DM clump. In conclusion,
being the two DM peaks offset with respect to the SZ$_{th}$ peaks, the kinetic SZ effect
is at most a marginal residual signal at the location of the two DM peaks of the cluster
1ES0657.\\
In these conditions the detectability of SZ$_{DM}$ is limited mainly by the fluctuations
of the unresolved galaxies background. Negative or positive fluctuations can be found at
the locations of the $SZ_{DM}$ peaks, affecting its detection. From realistic scan
simulations at 220 GHz with a $1 ^\prime$ FWHM beam, we estimate that a 2-$\sigma$
(3-$\sigma$) detection  has a likelihood $\simgt 95\%$ if the $SZ_{DM}$ in the beam is
$\simgt 6.5 \mu$K ($\simgt 10 \mu$K). These limits are comparable to the expected
$SZ_{DM}$ signal estimated in Sect 2.1. Complementary observations at higher and lower
frequencies should confirm the nature of the signal, which in the case of the unresolved
galaxies increases strongly with frequency, while is basically constant for the $SZ_{DM}$
There are several instruments expected to provide SZ surveys from the southern hemisphere
in the near future. The 10-m South Pole Telescope (SPT, Ruhl et al. 2004) will exploit
the excellent environmental conditions of the antarctic winter to produce deep surveys of
SZ clusters at 90, 150, 220 and 270 GHz. With a FWHM of $\simlt 1^\prime$ this instrument
is ideally suited to perform the DM search outlined here. The sensitivity of the SZ
survey will be $\sim 10 \mu K$ for each 1$^\prime$ pixel at 150 GHz: we suggest deeper
observations for 1E0657-558. A sensitivity of $\sim 0.5 \mu K$ per 1$^\prime$ pixel at
150 GHz can be obtained in 4 days of integration.
Increasing atmospheric and bolometer noise prevents reaching the same sensitivity in the
220 and 270 GHz bands (see also Gomez et al. 2003, Runyan et al. 2003 for the case of
experiments working at larger angular scales). Still, the sensitivity should be good
enough to see the two $\sim 10 \mu K$ negative signals of SZ$_{DM}$.
Similar considerations apply to the Atacama Cosmology Telescope (Kosowsky 2003), to the
APEX bolometer arrays (Lee et al. 2004) and to the BOLOCAM instrument at the CSO (Glenn
et al. 1998).
Fig.\ref{fig.sz_maps} shows the CMB maps expected from an observation of \es at $\nu =$
150, 223 and 350 GHz (with bandwidths of $\approx 25 \%$) with the SPT. The SZ$_{DM}$
effect clearly emerges localized at the two DM clumps for $\nu = 223$ GHz where SZ$_{th}$
vanishes for the brighter X-ray clump with $kT_e = 14$ keV. This is the optimal frequency
to detect the SZ$_{DM}$ from the cluster \es.
Such an experiment is able to set strong constraints on DM particle mass in the range
$M_{\chi} \sim 20-50$ GeV, while the SZ signals for $M_{\chi} \simgt 80$ GeV is
unobservable.
Higher frequency measurements are needed to monitor foreground contamination.
Observations of 1E0657-558 with the OLIMPO experiment (Masi et al. 2005), using bolometer
arrays at 145, 217, 345 and 545 GHz, with beam FWHM of 3.3$^\prime$, 2.2$^\prime$,
2.2$^\prime$, 2.2$^\prime$ respectively, will be able to resolve the two SZE peaks, and
will not be hampered by atmospheric fluctuations even in the highest frequency bands. In
10 hours of integration over the cluster area, the sensitivity of OLIMPO in the 217 (345)
GHz band will be $\approx$ 1.5 (3.0) $\mu$K  per pixel (with a pixel side of 1$^\prime$).
These measurements will nicely complement those of the ground based instruments, allowing
an effective separation of the different astrophysical components, with special
sensitivity to the fluctuating background from unresolved galaxies.
An exciting perspective is the upgrade of a space-borne telescope with a Fourier
Transform Spectrometer. This can provide the combination of high throughput and spectral
resolution needed to compare the SZ$_{th}$ null to the surrounding frequencies, thus
providing convincing evidence to distinguish between the different signal components (see
Fig.\ref{fig.sz_dm}). A first test can be carried out already with the OLIMPO experiment,
preparing the way for a future satellite mission.

\section{Conclusions}
The SZ$_{DM}$ effect is an inevitable consequence of the presence and of the nature of DM
in large-scale structures. Its analysis in the special case of \es can provide a direct
physical probe for the presence and for the nature of DM in cosmic structures.
\\
How these observations compare to other possible DM signal detections?
In general, both gamma-ray and radio observations of DM annihilation are quite powerful
to set constraints on the neutralino mass and composition (see, e.g. Colafrancesco et al.
2006, 2007).
However, in the case of \es  the expected gamma-ray emission associated to the DM clumps
is too low ($\simlt 1$ count vs. $\sim 10$ background counts at $E > 1$ GeV) and cannot
be resolved by GLAST from other possible sources of gamma-ray emission, both from the
cluster \es and from AGNs in the field. In addition, the GLAST spatial resolution ($\sim
9-18$ arcmin at 10 and 1 GeV, respectively) cannot provide any clear spatial separation
between the DM gamma-ray signals (expected to be concentrated at the DM clumps) and other
possible gamma-ray signals originating within the atmosphere of \es.
Radio telescopes have, in principle, excellent resolution and sensitivity to probe the
different spectra and brightness distribution of the DM-induced synchrotron emission.
For the sake of illustration, we evaluated that the DM-induced synchrotron emission from
the East DM clump is $\sim 3-10$ mJy (for a smooth or smooth plus $50 \%$ mass clumpiness
NFW DM profile, $M_{\chi}=40$ GeV, $\langle \sigma v \rangle_0 = 4.7 \cdot 10^{-25} cm^3
s^{-1}$ model used by Colafrancesco et al. 2006 for Coma, with a $1$ $\mu$G magnetic
field) at $\nu = 100$ MHz, still marginally detectable by LOFAR.
Theoretical uncertainties associated to the amplitude of the magnetic field in the DM
clumps of \es render, however, the prediction of the expected signals quite uncertain.\\
In such a context, the possible detection of the SZ$_{DM}$ effect will provide an
important complementary, and maybe unique, probe of the nature of DM.
\begin{acknowledgements}
The author thanks the Referee for useful comments and suggestions.
\end{acknowledgements}

\end{document}